\documentclass[%
reprint, 
superscriptaddress,
 amsmath,amssymb,
 aps,
 prx
]{revtex4-2}

\usepackage{braket}
\usepackage{graphicx}
\usepackage{dcolumn}
\usepackage{bm}
\usepackage{siunitx}
\usepackage{xcolor} 
\usepackage{tabularx}
\usepackage{dsfont}					
\usepackage{mathrsfs}				
\usepackage{float}
\usepackage{hyperref}

\newcommand{\tr}[0]{\mathbf{tr}}		

\newcommand{\be}[0]{\begin{equation}}	
\newcommand{\ee}[0]{\end{equation}}

\usepackage{hyperref}
\hypersetup{colorlinks, 
	linkcolor={blue!75!black!80!yellow},
	citecolor={blue!75!black!80!yellow}, 
	urlcolor={blue!75!black!80!yellow}
	}

\newcommand{\df}{\mathrm{d}}

\usepackage[normalem]{ulem}

\begin{document}


\title{Stabilizing fluctuating spin-triplet superconductivity in graphene via induced spin-orbit coupling}


\author{Jonathan B. Curtis}
\email{joncurtis@ucla.edu}
\affiliation{College of Letters and Science, University of California, Los Angeles, CA 90095, USA}
\affiliation{Department of Physics, Harvard University, Cambridge, MA 02138, USA}
\author{Nicholas R. Poniatowski}
\author{Yonglong Xie}
\author{Amir Yacoby}
\affiliation{Department of Physics, Harvard University, Cambridge, MA 02138, USA}
\author{Eugene Demler}
\affiliation{Institute for Theoretical Physics, ETH Z\"urich, 8093 Z\"urich, Switzerland}
\author{Prineha Narang}
\email{prineha@ucla.edu}
\affiliation{College of Letters and Science, University of California, Los Angeles, CA 90095, USA}


\date{\today}

\begin{abstract}
    A recent experiment showed that proximity induced Ising spin-orbit coupling enhances the spin-triplet superconductivity in Bernal bilayer graphene. Here, we show that, due to the nearly perfect spin rotation symmetry of graphene, the fluctuations of the spin orientation of the triplet order parameter suppress the superconducting transition to nearly zero temperature. 
    Our analysis shows that both Ising spin-orbit coupling and in-plane magnetic field can eliminate these low-lying fluctuations and can greatly enhance the transition temperature, consistent with the recent experiment. 
    Our model also suggests the possible existence of a phase at small anisotropy and magnetic field which exhibits quasi-long-range ordered spin-singlet charge 4e superconductivity, even while the triplet 2e superconducting order only exhibits short-ranged correlations. 
    Finally, we discuss relevant experimental signatures. 
\end{abstract}

\maketitle

\begin{figure}
    \centering
    \includegraphics[width=\linewidth]{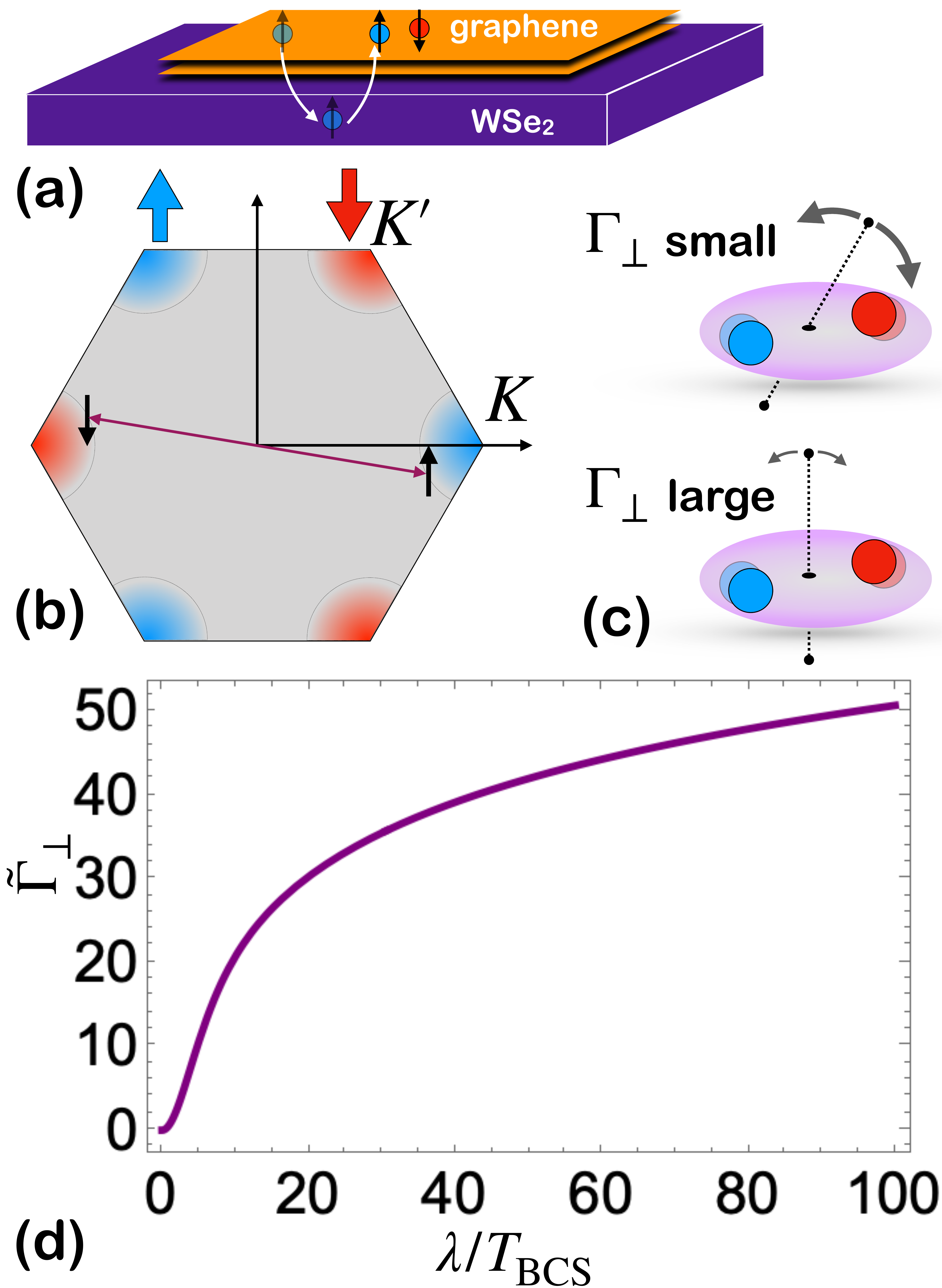}
    \caption{
    Schematic overview of system considered.
    (a) Illustration of Bernal bilayer graphene (BBG) on top of WSe$_2$. The WSe$_2$ proximity induces an Ising spin-orbit coupling in the graphene by virtual tunneling processes. 
    (b) Pairing in the graphene Brillouin zone, for the case of $f$-wave triplet pairing. 
    Pairing is between $K$ and $K'$ so that electrons are in Kramers pairs with total momentum zero while
    Ising spin-orbit coupling acts like a Zeeman field for spins pointing up at $K$ and down at $K'$.
    (c) Illustration of how spin-orbit coupling governs the scale of the d-vector pinning $\Gamma_\perp$.
    For small $\Gamma_\perp$ the fluctuations of the orientation of $\mathbf{d}$ are soft, whereas for large $\Gamma_\perp$ the $\mathbf{d}$-vector is strongly pinned along the $z$-axis.
    (d) BCS calculation of dimensionless pinning energy $\tilde{\Gamma}_\perp = \Gamma_\perp/N(E_F)$ as a function of the induced Ising spin-orbit coupling $\lambda$, relative to the BCS pairing temperature. 
   One can see that $\tilde{\Gamma}_\perp$ increases by over two orders of magnitude.}
    \label{fig:schematic}
\end{figure}

\textit{Introduction} \textemdash
Graphene based two-dimensional materials have emerged as a highly tunable platform for studying spin-triplet superconductivity. 
Notable examples include rhombohedral trilayer graphene (RTG)~\cite{ZhouABC2021}, moir{\'e} graphene~\cite{Cao2021,Park2022}, and Bernal bilayer graphene (BBG)~\cite{Zhou.2022k0q,Zhang.2022}.
In the case of BBG, superconductivity was initially found to emerge exclusively in the presence of an in-plane magnetic field that exceeds the paramagnetic limit, consistent with a spin triplet order parameter~\cite{Zhou.2022k0q}. 
A very recent experiment showed that introducing a tungsten diselenide (WSe$_2$) layer \textemdash a semiconducting material with strong spin-orbit coupling \textemdash on top of BBG, as schematically depicted in Fig.~\ref{fig:schematic}(a), can stabilize superconductivity at strictly zero magnetic field and can enhance the critical temperature by an order of magnitude~\cite{Zhang.2022}.

An important task prompted by this recent experiment is to identify the role of the proximity-induced spin-orbit coupling in promoting spin-triplet superconductivity in BBG. 
One possibility is that spin-orbit coupling selects a particular symmetry-broken parent state from which the superconductivity emerges, and previous work~\cite{Qin2022,You2022,Lu2022} shows that the intervalley coherent ground state is conducive to spin triplet pairing.
Alternatively, the presence of WSe$_2$ can introduce virtual tunneling process that enhances pair binding energies and hence boosts the critical temperature~\cite{Chou.2022}.

In this work, we present a novel mechanism by which the proximity-induced spin-orbit coupling promotes the spin-triplet superconductivity observed in graphene. 
Specifically, we show that the presence of Ising spin-orbit coupling breaks the nearly perfect spin-rotation symmetry in graphene and thus suppresses the order parameter fluctuations caused by the otherwise gapless Goldstone modes corresponding to fluctuations of the orientation (in spin space) of the triplet order parameter (for a precise definition of Ising-type spin-orbit coupling see the discussion below or Supplemental Material).
While triplet superconductivity is formally impossible in intrinsic graphene on account of the Mermin-Wagner theorem, it is allowed via a BKT-type transition once spin-rotation symmetry is broken by the WSe$_2$.
A similar proposal was recently put forth in Ref.~\cite{Cornfeld.2021}, motivated by the potential observations of triplet pairing in graphene.

To some extent, this phenomenology is relevant to all reports of spin-triplet superconductivity in graphene and may offer a degree of universality once one takes superconducting fluctuations in to account. 
In particular, we expect that all perturbations which break spin-rotation symmetry, such as in-plane magnetic field or induced Ising spin-orbit coupling, can help stabilize fluctuating spin-triplet superconductivity; a magnetic field stabilizes a spin-polarized condensate~\cite{Cornfeld.2021}, whereas Ising spin-orbit coupling stabilizes a unitary (spin nematic) triplet state.
By suppressing order-parameter fluctuations, these perturbations help restore the mean-field pairing temperature.
Our theory is based entirely on general symmetry arguments and is therefore broadly applicable to many graphene-based triplet superconductors including BBG, RTG, and moir{\'e} systems and highlights the important role of collective modes in designing spin-triplet superconductors with higher critical temperatures.

\textit{Microscopic Model}\textemdash We begin by considering a minimal microscopic model for BBG, consisting of two spin-degenerate parabolic bands at valleys $\mathbf{K}$ and $\mathbf{K}'$ in the presence of an induced Ising spin orbit coupling of strength $\lambda$. 
This is depicted in Fig.~\ref{fig:schematic}(b) which shows the intervalley pairing and induced spin-orbit coupling in the Brillouin zone. 

Introducing the four-component spinor $\psi_{\mathbf{k}}$ and Pauli matrices $\sigma_i$ ($\rho_i$) which act on the spin (valley) degrees of freedom, we can write the normal state Hamiltonian as $H = \sum_{\mathbf{k}} \psi_{\mathbf{k}}^\dag h_{\mathbf{k}} \psi_{\mathbf{k}}$ with $h_{\mathbf{k}} = \xi_{\mathbf{k}} \rho_0 \sigma_0 + \frac{1}{2} \lambda \rho_3 \sigma_3$ where $\xi_{\mathbf{k}}$ is the quadratic dispersion of each band. 
To include the effects of triplet pairing, we express the Hamiltonian in terms of the eight-component Nambu spinor $\Psi_{\mathbf{k}} = \left( \psi_{\mathbf{k}} \; \bar{\psi}_{-\mathbf{k}}i\sigma_2  \right)^T$ as $H = \frac{1}{2}\sum_{\mathbf{k}} \Psi^\dag_{\mathbf{k}} \hat{\mathcal{H}}_{\mathbf{k}} \Psi_{\mathbf{k}}$ with 
\begin{equation}
\label{eqn:hamiltonian}
    \hat{\mathcal{H}}_{\mathbf{k}} = \begin{pmatrix} \xi_{\mathbf{k}} + \frac{1}{2} \lambda \rho_3 \sigma_3 & \mathbf{d}\cdot \bm{\sigma} i\rho_2 \\ -\mathbf{d}^\star \cdot \bm{\sigma} i\rho_2 & -\xi_{\mathbf{k}} + \frac{1}{2}\lambda \rho_3 \sigma_3 \end{pmatrix} \, .
\end{equation}
This model features vector order parameter $\mathbf{d} = g_f \sum_{\mathbf{k}} \langle\psi_{\mathbf{k}}^T \left( \rho_2 \sigma_2 \bm{\sigma} \right) \psi_{-\mathbf{k}} \rangle$ which describes an isotropically gapped $f$-wave, valley-singlet, spin-triplet state such that the gap takes the opposite sign in the $\mathbf{K}$ and $\mathbf{K}'$ valleys~\footnote{The pairing is assumed isotropic in $\mathbf{k}$ with respect to the $\mathbf{K,K'}$ points, but is valley singlet meaning that under $C_3$ rotations the valleys exchange and the wavefunction is overall odd, making it $f$-wave.}. 

This spin-singlet, valley-triplet order parameter has recently been argued to generally correspond to the most energetically favorable state in graphene-based systems~\cite{chou-prl,Chou.PRB.2022,Lu2022}, but the analysis to follow should generalize to other triplet order parameters. 
Our crucial assumption is that in the absence of spin-orbit coupling the system has perfect $SU(2)$ symmetry, and that all subdominant pairing interactions are negligible such that, in particular, there are no valley collective modes (which unlike spin are not guaranteed to exist by symmetry).
We leave the case of specific broken symmetry normal states to future work.
This assumption allows us to project the spin-orbit coupling onto the triplet manifold while neglecting the mixing between singlet and triplet subspaces. 

Using this Hamiltonian, we can carry out the standard procedure for deriving the Ginzburg-Landau free-energy for the triplet order parameter $\mathbf{d}$ near the critical point (see Supplemental Material for details).
We work within the weak-coupling BCS limit~\footnote{In fact, it is likely that BBG is not in the weak coupling regime and this may lead to additional interesting effects, though the argument we advance is based on symmetry and is still a relevant consideration in such case.}, and find the free-energy density 
\begin{equation}
\label{eqn:fe}
    f = K |\nabla \mathbf{d}|^2 + r|\mathbf{d}|^2 + \Gamma_\perp |\mathbf{d}_{\perp}|^2+\frac12 u |\mathbf{d}|^4 + \frac12 v (i\mathbf{d} \times \mathbf{d}^\star )^2 .
\end{equation}
The first term with $K$ gives the superconducting phase stiffness, the second term is the bare pairing susceptibility, with $r \sim \log(T/T_{\rm BCS})$ changing sign at the mean-field BCS temperature, and $\Gamma_\perp$ is the spin-orbit induced pinning energy for the out-of-plane orientations, with $|\mathbf{d}_\perp|^2 = |d_x|^2 + |d_y|^2$ (see discussion below). 
The nonlinearity $u$ respects an enlarged $SU(3)$ symmetry while the nonlinearity $v = u$ further penalizes the non-unitary equal-spin pairing states, and breaks the symmetry down to $SO(3)$.

Crucially, for nonzero symmetry-breaking spin-orbit coupling $\lambda$ (see Hamiltonian in Eq.~\ref{eqn:hamiltonian} for definition) we find a finite $\mathbf{d}$-vector pinning energy $\Gamma_\perp$ which selects the $z$-axis as the favored orientation of the $\mathbf{d}$-vector, depicted in Fig.~\ref{fig:schematic}(c).
This pinning energy, $\Gamma_\perp$ is plotted as a function of $\lambda$ in Fig.~\ref{fig:schematic}(d). 
For small $\lambda/T_{\rm BCS}$ we find $\Gamma_\perp \sim \lambda^2/T_{\rm BCS}^2$, while for larger values the growth slows to a logarithmic scaling $\sim \log(\lambda)$. 
We note that the value of the induced spin-orbit coupling $\lambda \sim 0.7$ meV inferred from experiments \cite{Zhang.2022} is likely to be one of the larger energy scales in the problem and may possibly rival the Fermi energy of the small pockets, pointing to the need for a more detailed treatment in future work. 
For small $\Gamma_\perp$, the actual transition temperature $T_c$ will be suppressed due to strong fluctuations, as we now show. 



{\it Goldstone Modes}\textemdash
We now explore the phase diagram of the Ginzburg-Landau functional in Eq.~\eqref{eqn:fe} as a function of temperature $T$ and the pinning energy $\Gamma_\perp$. 
In fact, similar models have already been studied extensively in the context of BKT transitions in ultracold spinor condensates~\cite{Mukerjee.2006,Podolsky.2009,Kobayashi.2019}, where Monte Carlo studies have been conducted for a range of possible anisotropy and interaction parameters~\cite{Kobayashi.2019}.
In this work, we will focus on the qualitative effect that Gaussian fluctuations have on the phase diagram, rather than quantitatively determine the phase diagram since there is anyways still great debate about the model and parameters, which can also vary from system to system. 

We treat this system using a large-$N$ expansion in the number of spin components, as detailed in the Supplemental Material.
Since this is a Gaussian approximation, it does not capture topological defects such as vortices and is expected to underestimate the role of fluctuations.
This approach does not allow us to distinguish between long-range ordered phases and quasi-long-range ordered (QLRO) phases.
However, it is known that in two-dimensions at finite temperature fluctuations of the superconducting phase do not permit true long-range order, and thus when we discuss the symmetry breaking phases with $\langle d_z\rangle \neq 0$ it is understood that these are in reality QLRO. 
Nevertheless, this approach allows us to study the impact of spin-orbit coupling on fluctuations in a controlled way by finding the renormalized critical temperature $T_c$ where the $\langle d_z\rangle$ QLRO condensate develops.
We also find a small region of $4e$ QLRO superconductivity at small anisotropy which will be discussed below.

The key quantity of interest is the condensate $\psi \sim \langle d_z \rangle$. 
Within mean-field theory, this is simply the solution to the saddle-point equation and is given by $|\psi| = \sqrt{-r/u}$ when $T < T_{\rm BCS}$ and zero above. 
Once the order-parameter fluctuations are taken in to account~\cite{Larkin.2005,Varlamov.1999}, we find the renormalized critical temperature $T_c$ given by 
\begin{equation}
    \log\left(\frac{T_c}{T_{\rm BCS}}\right) = -{\rm Gi}_{(2)} \log\left(\frac{C}{\tilde{\Gamma}_\perp}\right).
\end{equation}
Here $\tilde{\Gamma}_\perp = \Gamma_\perp/N(E_F)$ is the unitless pinning energy relative to the density-of-states at the Fermi level, $C$ is a unitless ultraviolet cutoff on fluctuations given by $C = K q_{\rm max}^2/N(E_F) \sim E_F^2/T_{\rm BCS}^2\sim 1/\textrm{Gi}_{(2)}^2$. 
The important parameter here is the Ginzburg-Levanyuk~\cite{Larkin.2005,Varlamov.1999} number
\begin{equation}
    \textrm{Gi}_{(2)} = T_{\rm BCS}\frac{u}{2 \pi K N(E_F)} = \frac{T_{\rm BCS}}{\pi v_F^2 N(E_F)} =\frac{T_{\rm BCS}}{E_F},
\end{equation}
which is a measure of the strength of superconducting fluctuations. 

Solving this for $T_c$ we find a power-law behavior in $\Gamma_\perp$ scaling as $T_c \sim \Gamma_\perp^{\textrm{Gi}_{(2)}}$. 
The Ginzburg-Levanyuk parameter is not known well experimentally in this system since it requires a knowledge of the ``pseudogap" temperature characterized by a spectral gap appearing at $T_{\rm BCS}$, and spectroscopic experiments have not yet been performed on this material.
In other graphene-based superconductors however, it has been argued that this may be quite large with values of $T_{\rm BCS}/T_F \sim 0.1$~\cite{Park.2021}. 
Tunneling spectroscopy measurements have also found evidence for a substantial quasiparticle pseudogap~\cite{Oh.2021,Kim2022}.
Furthermore, superconductivity in BBG occurs proximal to different Fermi surface topologies which may feature small Fermi surfaces with small Fermi energies, which may only be as small as $0.6$ meV (for details see Supplemental Material).
We will roughly estimate a pairing temperature of $T_{\rm BCS} \sim 0.5$ K~\footnote{While this may be somewhat optimistic, closer examination of the data in Ref.~\cite{Zhang.2022}, and in particular the inset of Fig. 2(d), reveals a possible Azlamazov-Larkin downturn in $R_{xx}$ around $T\sim 500$ mK, indicating this may be a serious possibility.}, and a Fermi temperature of $T_F \sim 0.6\; \textrm{meV}\sim 7$ K, in line with strong-coupling, with $\textrm{Gi}_{(2)} \sim 0.07$.
Intrinsic spin-orbit coupling in graphene is expected to be no more than 40 $\mu$eV or so~\cite{Sichau.2019}, and once placed on top of WSe$_2$ is estimated to be of order $0.7$ meV, corresponding to at least a 20-fold increase in $\lambda$, meaning an increase in ${\Gamma}_\perp$ of order $60$-fold is feasible~\footnote{This is a lower bound on the increase. In particular, it is thought that the intrinsic spin-orbit coupling in graphene takes the form of $H_{\rm soc} = \lambda \sigma_3 \rho_3 s_3$ where $s_a$ are the sublattice Pauli matrices. Thus, in AB stacked graphene the average spin-orbit coupling still vanishes and the intrinsic value is likely much less than 40 $\si{\micro\eV}$ to begin with.}.
This could increase $T_c$ by a factor of 33\%, however since we have only gone to leading order in $1/N$ (recall $N = 3$), our approximations likely underestimate the effect of fluctuations and give a {\it conservative} estimate.

\begin{figure}
    \centering
    \includegraphics[width=\linewidth]{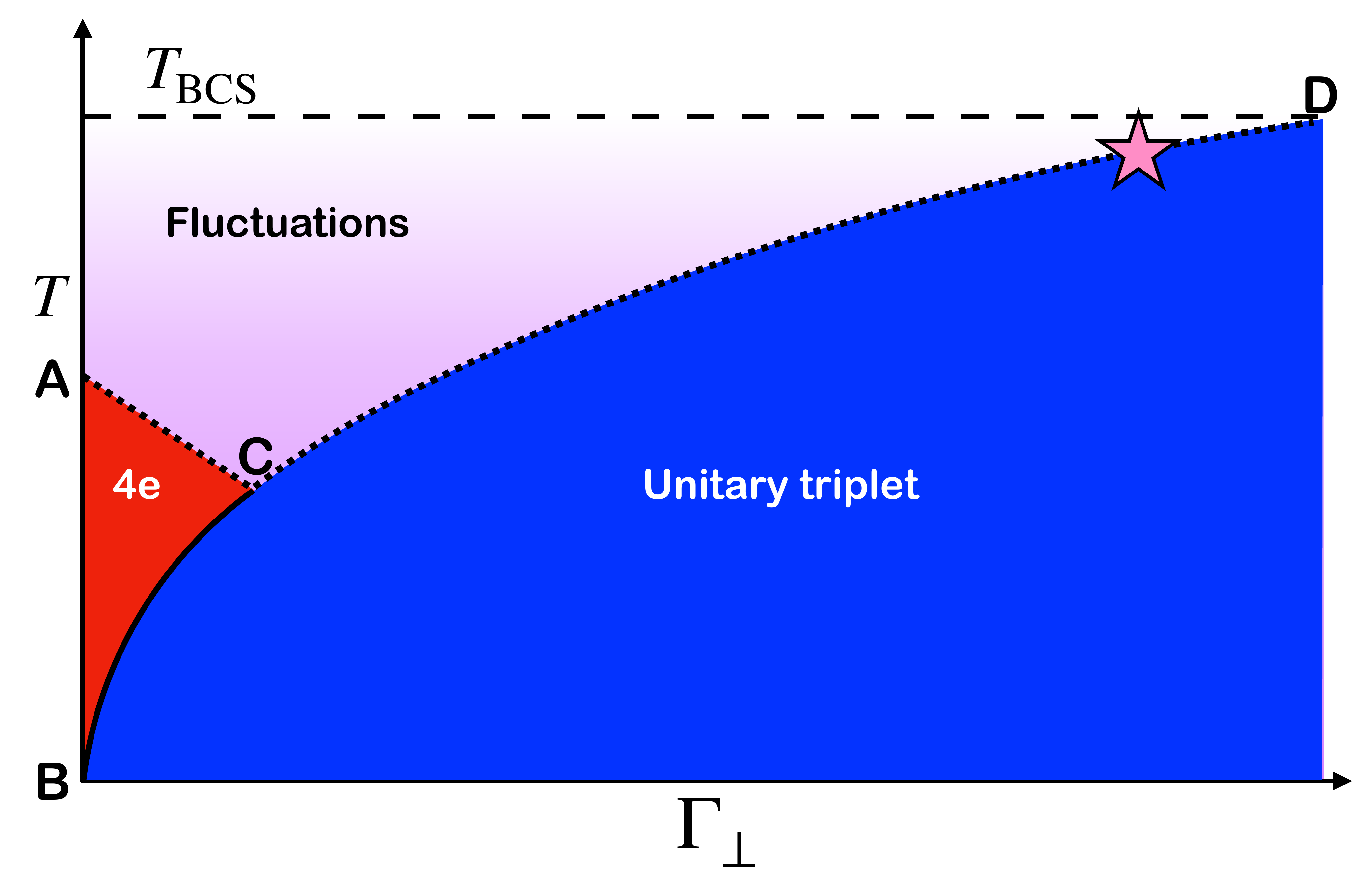}
    \caption{
    Phase diagram of triplet superconductivity in graphene in the $T-\Gamma_\perp$ plane.
    For temperatures below the mean-field BCS transition $T_{\rm BCS}$ (dashed line) there are strong superconducting fluctuations which are especially pronounced in the isotropic triplet system.
    Ising spin-orbit coupling $\Gamma_\perp$ pins the $\mathbf{d}$ vector and induces a gap for the unitary Goldstone modes, suppressing fluctuations and enhancing the renormalized transition into a unitary triplet superconducting state with $\langle d_z\rangle$ pairing.
    This leads to a quasi-long-range ordered unitary triplet superconducting phase (blue region) below the renormalized critical temperature (dashed line {\bf C-D}) via BKT transition. 
    At small spin-orbit coupling the Goldstone modes strongly suppress the triplet transition, but allow for a quasi-long-range ordered singlet $4e$ superconductor (dashed line {\bf A-C}), also via a BKT transition.
    In principle, at lower temperatures in the $4e$ phase there may be a second Ising transition which spontaneously breaks the residual $\mathbb{Z}_2$ spin-symmetry of $d_z\to -d_z$ (solid line {\bf B-C}), terminating in a tricritical point.
    In the presence of a substrate with large spin-orbit coupling, such as the WSe$_2$, we reside at the star shown, where the transition is essentially restored to $T_{\rm BCS}$. 
    The size of the QLRO $4e$ phase is over-exaggerated for clarity in this diagram.}
    \label{fig:phase-diagram}
\end{figure}

This behavior is illustrated schematically in Fig.~\ref{fig:phase-diagram}.
We briefly mention that this is the phase diagram {\it before} including the topological vortex defects, which are responsible for the BKT transition known to occur in all two-dimensional superfluids. 
We qualitatively expect that this will lead to a further renormalization of the boundary of the superfluid phase~\cite{Kobayashi.2019} with the relevant BKT temperature based on the topological classification of the defects. 
Our estimates therefore indicate that at least some part of the observed spin-orbit induced enhancement of $T_c$ may be due to suppression of superconducting fluctuations. 

{\it Magnetic Field}\textemdash
In addition to proximity inducing spin-orbit coupling, we briefly consider breaking the spin-rotation symmetry by applying an in-plane magnetic field, as was done in the Santa Barbara experiments which first saw putative triplet pairing in BBG~\cite{Zhou.2022k0q}. 
The relevant term in the Ginzburg-Landau expansion is not the $\Gamma |\mathbf{d}_\perp|^2$ term but rather the term 
\begin{equation}
    f_{\rm Zeeman} = -\mu_{\rm eff}\mathbf{H}_{\parallel}\cdot( i\mathbf{d} \times \mathbf{d}^\star ),
\end{equation}
where $\mathbf{H}_{\parallel} = H_{\rm ext} \mathbf{e}_x$ is the applied in-plane magnetic field that couples to the non-unitary condensate moment, which has an effective magnetic moment of $\mu_{\rm eff}$.

While we find that within the BCS model and quasiclassical approximation this coupling vanishes, we expect that in a strongly-coupled superconductor this coupling will in general appear with a coefficient proportional to the degree of particle-hole symmetry breaking; in the Supplemental Material we find $\mu_{\rm eff} \sim 2\mu_B N'(E_F)/\tilde{g}_f$ where $N'(E_F)$ is a measure of the particle-hole asymmetry and $\tilde{g}_f = N(E_F)g_f$ is the dimensionless BCS coupling constant in the relevant channel.
We estimate that for a $100$ mT field, $\mu_{\rm eff}H_{\rm ext} \sim \textrm{Gi}_{(2)}\times \frac{2\mu_B H_{\rm ext}}{T_{\rm BCS}}\times\frac{N(E_F)}{\tilde{g}_{f}} \sim 0.2\, N(E_F)$, assuming a BCS coupling constant of $\tilde{g}_f \sim 0.1$~\footnote{This is to be distinguished from the ratio of $T_{\rm BCS}/E_F$, though in general it is a conceptual question how to characterize the strength of the pairing interaction in the regime of large Ginzburg number.}.

At the level of mean-field theory, this term has two effects.
The first is that it actually slightly raises $T_c$, and the second is that it causes the transition to now enter into a fully spin-polarized condensate, analogous to the A$_1$ phase of $^3$He~\cite{Kojima.2008}. 
To see this, we can change basis to the Zeeman sublevels $m=1,0,-1$ of the pair spin along the direction of the in-plane field, with condensates $d_{m}$. 
The free energy up to quadratic order is then characterized by the potential 
\begin{equation}
    V(d_{+1}) = (r+H_{\parallel}) |d_{-1}|^2 + r |d_0|^2 + (r - H_{\parallel})|d_{+1}|^2.
\end{equation}
Evidently, the system first orders into the $m=1$ condensate once $r - H_{\parallel}$ becomes zero.
This gain in energy is linear in the condensate density, whereas the energy penalty for non-unitary pairing due to the $v$ coefficient is only quadratic and thus near $T_c$, when the condensate density is small, it will always enter into a fully-polarized condensate.

In the polarized phase the magnetic field is expected to also suppress fluctuations since it breaks spin-rotation symmetry.
One can check that there is only one remaining Goldstone mode, which is the overall $U(1)$ gauge symmetry; the remaining modes are pinned by the field with a gap of order $|\mu_{\rm eff}H_{\rm ext}|$.
While a quantitative treatment of this effect is left to future studies, one would then expect a similar behavior as to the case of the Ising spin-orbit coupling, with the replacement of $\Gamma_\perp \to \mu_{\rm eff} H_{\rm ext}$.
Comparing the pinning energy of the lowest-lying Goldstone mode between the case of magnetic field and spin-orbit coupling, we see that the spin-orbit interaction, with $\Gamma_\perp \sim 30 \,N(E_F)$, is roughly $150$ times stronger than the pinning energy of the 100 mT field, which gives $gH_{\rm ext} \sim 0.2 \, N(E_F)$. 
It is therefore plausible that there will still be a significant decrease in the fluctuations when passing from a 100 mT magnetic field to the 0.7 meV spin-orbit interaction.

{\it 4e-Condensate}\textemdash
We now return to the existence of $4e$-pairing.
In the ansatz used to solve the large-$N$ theory, this corresponds to a complex order parameter $\chi$, which couples to $\mathbf{d}^2$.
This parameter allows for a breaking of the $U(1)$ gauge symmetry while still leaving the $SO(3)$ spin-rotation symmetry unbroken, with a condensate of $\langle \mathbf{d}^2\rangle\neq 0$. 
Therefore, this $4e$ condensate is more robust against fluctuations and can in principle appear at small spin-orbit coupling $\Gamma_\perp$. 

This is seen in the phase-diagram (greatly exaggerated relative to the mean-field condensate) in Fig.~\ref{fig:phase-diagram}.
The boundary of this phase is also determined in the Supplemental Material by minimizing the variational ansatz. 
We find the $4e$ condensate appears at a temperature $T^\star$ found by solving
\begin{equation}
    \log\left(\frac{T^\star}{T_{\rm BCS}}\right) = \frac12\textrm{Gi}_{(2)} - \tilde{\Gamma}_\perp -\textrm{Gi}_{(2)} \log\left(\frac{2C}{\textrm{Gi}_{(2)}}\right).
\end{equation}
Thus, the phase boundary decreases exponentially with increasing $\Gamma_\perp$ before eventually colliding with the phase boundary of $2e$ condensate at a multi-critical point, with critical spin-orbit coupling $\Gamma_{\perp c} = N(E_F) \textrm{Gi}_{(2)} $. 

This phase can be thought of as a fluctuating superfluid of triplet pairs which exhibit correlations in the phase of their fluctuations while remaining randomly oriented in their spin orientation. 
$4e$ superconductivity has been proposed recentlyin graphene based systems~\cite{Maccari.2022,Khalaf.2022,Fernandes.2021} as well as quantum gases~\cite{Mukerjee.2006,Kobayashi.2019}, and would be accompanied by a characteristic flux quantum of $h/4e$ \textemdash half the usual value. 

Further lowering the temperature, we expect that the $4e$ phase will exhibit a second phase transition into the unitary (spin nematic) phase characterized by a condensate in $\langle d_z\rangle$, shown in Fig.~\ref{fig:phase-diagram} as the blue region at lower temperatures below the red region. 
Because the $4e$ condensate has half the standard flux quantum, it only fixes the phase of the condensate modulo $\pi$ \textemdash alternatively, this is understood as a Josephson coupling between the $4e$ condensate $\chi$ and the $2e$ condensate of the form $f_{\rm int} \sim \overline{\chi}\psi^2 + \textrm{c.c.}$ and thus the phase of $\psi$ still has a remnant $\mathbb{Z}_2$ symmetry which is broken at the second transition.
It is worth mentioning that this phase is in reality quite subtle~\cite{Kobayashi.2019} and may be overwhelmed by competing orders such as the intervalley coherent order~\cite{Chatterjee.2021a7j}

{\it Experimental Signatures}
It is likely possible to tune the strength of the induced Ising spin-orbit coupling, and hence the pinning energy $\Gamma_\perp$, by adding spacer layers of an inert material like hexagonal boron-nitride or even an oxide barrier in order to determine the functional dependence of $T_c$ on $\Gamma_\perp$.
This may also be possible to vary by tuning the twist-angle between the BBG and WSe$_2$~\cite{Chou.2022}.
Our predictions also should carry over to other semiconductors besides WSe$_2$ so long as they feature large Ising spin-orbit coupling and interface well with graphene. 

As to signatures of the fluctuations, a good starting point may be measurements of the Nernst-effect, which is sensitive to mobile vortices forming above $T_c$~\cite{Ussishkin.2002,Cyr.2018,Levchenko.2011,Jotzu.2021}.
It may also be possible to image the spin-fluctuations directly using scanning SQUID~\cite{Vasyukov2013} or NV magnetometry~\cite{Dolgirev.2022,Chatterjee.2022}, though this is complicated somewhat by the unitary nature of the fluctuations, which have no net magnetic dipole moment. 

In general, we expect that the presence of strong triplet fluctuations will lead to a strongly non-BCS character of the spin-susceptibility in the normal state just above $T_c$.
This is explored in more detail in the Supplemental Material, but neglecting the $4e$ pairing, we expect that the zero-field in-plane magnetic susceptibility (which probes only the spin susceptbility) has a characteristic behavior of $\chi_\parallel \sim \textrm{Gi}_{(2)}/(r_{\rm eff} + \Gamma_\perp)$ and for small spin-orbit coupling this has an apparent divergence as the critical point $r_{\rm eff}=0$ is approached, though this may be complicated by the emergence of $4e$-singlet or other competing orders at small spin-orbit coupling. 
We note this is in contrast to the standard case of a BCS triplet superconductor, which has no anomalous behavior at the critical point in the transverse spin-susceptibility. 

Finally, spectroscopic measurements would be useful in establishing the existence of a pseudogap phase above the actual transition, which in turn would indicate the presence of strong-fluctuations. 
This has already been accomplished in the case of twisted bilayer and trilayer graphene, which have indicated at least strong-coupling physics~\cite{Park.2021,Oh.2021}. 
Potentially, there may also be spectroscopic signatures of the collective modes themselves which would be interesting to probe~\cite{Poniatowski.2021zr,Curtis.2022,Poniatowski.2022,Dolgirev.2021ouj}. 
In particular, signatures of dipolar condensate waves~\cite{Poniatowski.2021zr}, or degenerate Bardasis-Schrieffer modes~\cite{Curtis.2022} should point towards triplet pairing in this system.

{\it Conclusion}\textemdash
In conclusion, we have presented evidence which supports the claim that graphene multilayers, and in particular BBG, may support triplet superconductivity.
In its intrinsic form, graphene has almost perfect spin-rotation symmetry and this, combined with the two-dimensional nature of graphene, leads to remarkably strong fluctuation effects which profoundly affect the phase diagram. 
Applying either Ising spin-orbit coupling or an in-plane Zeeman field would suppress these fluctuations and lead to an enhancement of the superconducting transition temperature, as have been experimentally observed.  

While our work is to some degree universally applicable to triplet pairing in graphene, there remain a number of qualitative complications due to the particular broken symmetry states which are proximal to the triplet pairing. 
It therefore remains important to understand what role the fermiology plays and how this couples to the fluctuation physics we have outlined here.
In particular, it is interesting to consider the interplay between the fluctuating triplet order and possible charge-density wave orders, such as the intervalley coherent order~\cite{Chatterjee.2021a7j,Zhou.2022k0q,Zhang.2022}. 
Treating the joint fluctuations of these two order parameters is a very interesting problem which may help explain certain aspects of the system at low magnetic field.
One can imagine that by expanding the symmetry group from $SO(3)\times U(1)$ to include other symmetries, one can obtain a unified~\cite{Zhang.1997,Podolsky.2004,Murakami.1999} model of both the isospin ordered and triplet paired states, like in Refs.~\cite{Khalaf.2021r9j,Christos.2020}.
Similarly, recent proposals for a spontaneous Ising spin-orbit coupling~\cite{Lake.2022} may also naturally fall out of a higher-symmetry parent phase which spontaneously breaks $SO(3)$ before $U(1)$ (and thus should also have soft Goldstone modes associated to the precession of the spontaneous Ising axis).

In the future, more direct evidence of triplet pairing, and of fluctuating superconductivity with a pseudogap, is still needed. 
Theoretically, it will be important in the future to account for the role of broken symmetry normal states and their potentially competing order parameters in our model.
Similarly, more careful consideration of the role of vortices and BKT physics is also needed, especially to make contact with experiments.
We believe that generally, the study of fluctuation effects in low-dimensional, clean, highly-symmetric graphene systems provides a unique opportunity to study superconductivity beyond mean-field, and in the presence of strong correlations.

\begin{acknowledgements}
The authors would like to acknowledge invaluable discussions with {\'E}tienne Lantagne-Hurtubise, Andrea Young, Ludwig Holleis, Pavel Dolgirev, Ilya Esterlis, and Andrew Zimmerman, and Olivia Liebman. J.B.C., N.R.P., Y.X., A.Y., and P.N. are supported by the Quantum Science Center (QSC), a National Quantum Information Science Research Center of  the  U.S.  Department  of  Energy  (DOE). J.B.C. and Y.X. are HQI Prize Postdoctoral Fellows and gratefully acknowledge support from the Harvard Quantum Initiative. N.R.P. is supported by the Army Research Office through an NDSEG fellowship. E.D. acknowledges support from the ETH Z\"urich Institute for Theoretical Physics. 
P.N. is a Moore Inventor Fellow and gratefully acknowledges support through Grant GBMF8048 from the Gordon and Betty Moore Foundation. 

\end{acknowledgements}

\bibliography{refs-1,refs-2,refs-yonglong}


\clearpage
\onecolumngrid
\appendix*


\section{Microscopic BCS Model}
In this section, we derive the Ginzburg-Landau free energy functional from the microscopic weak coupling theory. As introduced in the main text, we consider a two-dimensional system subject to an Ising spin-orbit coupling of strength $\lambda$, and an attractive interaction in the spin-triplet valley-singlet channel, with the coupling constant $g_f$. We treat the system within the functional integral formalism in the the Matsubara representation, and introduce the four-momenta $q = (i\omega_m,\mathbf{q})$ and $k = (i\epsilon_n,\mathbf{k})$ where $i\omega_m$ ($i\epsilon_n$) is a bosonic (fermionic) Matsubara frequency. The action governing the system is then written as 
\begin{equation}
\label{eqn:action}
    S = \sum_k \psi_{k}^\dag \left(-i\epsilon_n + \xi_{\mathbf{k}} + \frac{\lambda}{2} \, \sigma_3 \rho_3 \right) \psi_k - \frac{g_f}{2} \sum_q \mathbf{P}_q^\dag \mathbf{P}_q 
\end{equation}
where $\psi_k$ is a four-component spinor in spin $\otimes$ valley space and $\sigma_i$ ($\rho_i$) are the Pauli matrices acting on the spin (valley) degrees of freedom. We take the single-particle dispersion $\xi_{\mathbf{k}}$ to be parabolic in each of the valleys, and have defined the pair bilinear
\begin{equation}
    \mathbf{P}_q^\dag = \sum_k \bar{\psi}^T_{k+q/2} \left(i\bm{\sigma}\sigma_2 \otimes i\rho_2 \right) \bar{\psi}_{-k+q/2} \, .
\end{equation}
We introduce the vector superconducting order parameter $\mathbf{d}_q$ via a Hubbard-Stratonovich transformation. We express the action by doubling our basis and defining the eight-component Nambu spinor in spin $\otimes$ valley $\otimes$ particle-hole space, $\Psi_k = \left( \psi_k \, ,  \bar{\psi}_{-k} i\sigma_2 \ \right)^T$, so that the action takes the form
\begin{equation}
    S = \frac{1}{2g_f}\sum_q \bar{\mathbf{d}}_q \cdot \mathbf{d}_q - \frac{1}{2}\sum_{kq} \Psi_{k+q/2}^\dag \left[ \check{\mathbb{G}}^{-1}_k - \check{\Delta}_q  \right] \Psi_{k-q/2}
\end{equation}
where the normal-state Green function and pairing vertex are
\begin{align}
    \check{\mathbb{G}}^{-1}_k &= i\epsilon_n - \xi_{\mathbf{k}} \tau_3  - \frac{\lambda}{2} \sigma_3 \rho_3 \, ,\\[5pt]
    \check{\Delta}_q &= \mathbf{d}_q \cdot \bm{\sigma} (i\rho_2) \, \tau^+ + \bar{\mathbf{d}}_{-q} \cdot \bm{\sigma} (-i\rho_2) \, \tau^- \, .
\end{align}
In the above, we have introduced the Pauli matrices $\tau_i$ in Nambu space, as well as the combinations $\tau^{\pm} = \frac{1}{2} \left(\tau_1 \pm i\tau_2 \right)$. We may now perform the Gaussian integral over the fermion fields to furnish an effective action for the order parameter field, 
\begin{equation}
    S = \frac{1}{2g_f}\sum_q \bar{\mathbf{d}}_q \cdot \mathbf{d}_q - \frac{1}{2} \tr \log \left[ \check{\mathbb{G}}^{-1}_k - \check{\Delta}_q \right] \, .
\end{equation}
In the vicinity of the superconducting transition where the order parameter is small, we can expand the action in powers of $\mathbf{d}$ which, in the static $i\omega_n = 0$ limit, amounts to a derivation of the Ginzburg-Landau functional studied in the main text. 

To begin, we expand the functional logarithm to quadratic order, yielding 
\begin{equation}
    S^{(2)} = \frac{1}{2g_f}\sum_q \bar{\mathbf{d}}_q \cdot \mathbf{d}_q + \frac{T}{4}\sum_{kq} \tr \left[ \check{\mathbb{G}}_{k+q/2} \check{\Delta}_q \check{\mathbb{G}}_{k-q/2} \check{\Delta}_{-q}  \right] \, .
\end{equation}
The Ising spin-orbit coupling breaks spin rotation symmetry and leads to different propagators for the $d_z$ and $d_x, d_y$ fluctuations. The quadratic action can then be expressed as 
\begin{equation}
    S^{(2)} = \frac{1}{2}\sum_q \left[ \mathcal{L}^{-1}_{zz}(0,\mathbf{q}) \bar{d}^z_q d^z_q + \mathcal{L}^{-1}_\perp(0,\mathbf{q}) \left( \bar{d}^x_q d_q^x + \bar{d}^y_q d_q^y \right)   \right] \, ,
\end{equation}
The fluctuation propogators can be further expanded for small $\mathbf{q}$ as
\begin{align}
    \mathcal{L}_{zz}^{-1}(0,\mathbf{q}) &= r + K_{zz} \mathbf{q}^2 + \dots \\
    \mathcal{L}_{\perp}^{-1}(0,\mathbf{q}) &= r + \Gamma + K_\perp \mathbf{q}^2 + \dots \, .
\end{align}
which allows us to compute the quadratic coefficients appearing in the Ginzburg-Landau functional studied in the main text. In particular, we find that the $r$ coefficient reflects the singular piece of the fluctuation propogator that determines the mean field transition temperature $T_{BCS} = (4\mathrm{e}^{\gamma}/\pi) \, \mathrm{e}^{-1/N(E_F) g_f}$, 
\begin{equation}
    r = \mathcal{L}^{-1}_{zz}(0,0) = \frac{1}{g_f} - 2T\sum_k \left[\frac{1}{\epsilon_n^2 + (\xi_{\mathbf{k}} +\lambda/2)^2 } + \frac{1}{\epsilon_n^2 + (\xi_{\mathbf{k}} -\lambda/2)^2 } \right] = \nu \log \left( \frac{T}{T_{BCS}} \right) \, .
\end{equation}
Meanwhile, the $\mathbf{d}$-vector pinning is
\begin{equation}
\Gamma = \mathcal{L}^{-1}_\perp(0,0) - \mathcal{L}_{zz}^{-1}(0,0) = 2T\sum_k \frac{\lambda^2}{ [\epsilon_n^2 + (\xi_{\mathbf{k}} + \lambda/2)^2] [\epsilon_n^2 + (\xi_{\mathbf{k}} - \lambda/2)^2] } \, .
\end{equation}
This quantity is plotted as function of the spin orbit strength $\lambda$ in Fig. 1 of the main text, evaluated in the approximation that the density of states is constant over the energy range relevant for superconductivity, i.e. $N(\xi_{\mathbf{k}}) \approx N(E_F)$. 

For simplicity, we evaluate the coefficients of the gradient terms $K_{zz} = K_\perp \equiv K$ to lowest order in $\lambda$. We also expand the action to fourth order for $\mathbf{q}=0$, and again evaluate the resulting expansion coefficients to leading order in $\lambda$. The results for these coefficients are presented in Table \ref{tab:gl-coefficients}.

\begin{table}
\begin{center}
\begin{tabular}{ |c|c| }
 \hline
 GL Parameter & Weak Coupling Value  \\ 
 \hline 
 $r$ & $N(E_F)\log(T/T_{\rm BCS})$   \\
 $u$ & $7 \zeta(3) N(E_F)/(8 \pi^2 T_{\rm BCS}^2)$ \\
 $v$ & $7 \zeta(3) N(E_F)/(8 \pi^2 T_{\rm BCS}^2)$ \\
 $\Gamma_\perp$ & $7\zeta(3) N(E_F)\lambda^2/(4 \pi^2 T_{\rm BCS}^2)$ \\
 $K$ & $7\zeta(3)N(E_F) v_F^2/(16 \pi^2 T_{\rm BCS}^2)$ \\
 \hline
\end{tabular}
\end{center}
\caption{Coefficients appearing in Ginzburg-Landau functional and their weak coupling values, evaluated to leading order in the induced spin orbit coupling, $\lambda$.}
\label{tab:gl-coefficients}
\end{table}

\section{Zeeman field}
In this section, we consider the Ginzburg-Landau functional for a system subject to an in-plane magnetic field, $H_\parallel$, in the absence of spin-orbit coupling. The technical details of the calculation are identical to the previous section, with the replacement of the normal-state Green function by
\begin{equation}
    \check{\mathbb{G}}^{-1}_k = i\epsilon_n - \xi_{\mathbf{k}}\tau_3 - H_\parallel \sigma_3 \tau_3 \, .
\end{equation}
By computing the quadratic action, a new term appears which corresponds to a linear coupling at $\mathbf{q} = 0$ between the magnetic field and the condensate polarization,
\begin{equation}
    S_{\text{lin}} = -\mu_{\text{eff}} \sum_q H_\parallel \left( i \mathbf{d}_q \times \bar{\mathbf{d}}_q \right)_z \, ,
\end{equation}
where the coupling constant is 
\begin{equation}
    \mu_{\text{eff}} = 8T\sum_k \frac{\xi_{\bf k}}{[(i\epsilon_n - \xi_{\mathbf{k}})^2 - H_\parallel^2][(i\epsilon_n + \xi_{\mathbf{k}})^2 - H_\parallel^2]  } \, .
\end{equation}
To capture the linear response to weak fields, we can evaluate this coupling constant for $H_\parallel = 0$, and express the sum over momenta as an integral over energy weighted by the density of states $N(\xi)$, so that
\begin{equation}
    \mu_{\text{eff}} = 8T \sum_{\epsilon_n} \int \df \xi \; N(\xi) \, \frac{\xi}{(\epsilon_n^2 + \xi^2)^2}  \, .
\end{equation}
This expression is odd under $\xi \to -\xi$, and thus vanishes in the usual quasiclassical approximation $N(\xi) \approx N(E_F)$ which treats the approximate particle-hole symmetry of the weak coupling theory as exact. Thus, we see that the coupling between the condensate polarization and field is proportional to the degree of particle-hole asymmetry. Taking this to be small, we may then expand the density of states around its value at the Fermi level and write $N(\xi) \approx N(E_F) + N'(E_F) \,\xi$, which generates the leading order contribution to $\mu_{\text{eff}}$,
\begin{equation}
   \mu_{\text{eff}} = 4\pi T N'(E_F) \sum_{\epsilon_n} \frac{1}{\epsilon_n} \approx 4 N'(E_F) \left[\frac{1}{N(E_F) g_f} +  \log \left( \frac{T_{BCS}}{T}  \right) \right]\, .
\end{equation}

\section{Large-$N$ Expansion}
Here we approach the problem of order-parameter fluctuations via a $1/N$ expansion in the number of spin components, which in reality is $N=3$.
But as we all know, $3 = \infty$.
Of course, in reality this means that the effect of fluctuations is likely underestimated in this approach, and furthermore only Gaussian fluctuations are accounted for so that topological textures like vortices and half-vortices are also unaccounted for.

This approach will only be applied to the case of the Ising spin-orbit coupling, since in this case it is clear how to generalize to the $N$-component spinor, but in the case of magnetic field it is less clear. 
The appropriate generalization to the $N$-component unitary triplet pairing is found by writing $(i\mathbf{d}^*\times\mathbf{d})^2 = |\mathbf{d}|^4 - |\mathbf{d}\cdot\mathbf{d}|^2$, which can be generalized beyond the vector representation.

We then have 
\begin{equation}
    f = K |\nabla \mathbf{d}|^2 + r|\mathbf{d}|^2 + \Gamma_\perp |\mathbf{d}_\perp|^2 + \frac{(u+v)}{2N}|\mathbf{d}|^4 - \frac{v}{2N}|\mathbf{d}\cdot\mathbf{d}|^2 ,
\end{equation}
where the separation into $\perp$ and $\parallel$ spin components is selected by the Ising spin-orbit $\Gamma_\perp$.

We now solve this through a Gaussian variational ansatz for the free energy, evaluated in the large $N$ limit. 
The ansatz we use is based on the Feynman-Bogoliubov-Gibbs inequality such that 
\begin{equation}
    \mathcal{F} \leq \mathcal{F}_{\rm eff} = \mathcal{F}_0 + \langle \Delta \mathcal{F} \rangle_{\mathcal{F}_0}
\end{equation}
where $\mathcal{F}_0$ is the free energy of the ansatz and $\Delta \mathcal{F} = \mathcal{F} - \mathcal{F}_0$.
If we choose $\mathcal{F}_0$ to be Gaussian then this can be evaluated easily. 
By minimizing over our ansatz parameters we can best approximate the true free energy $\mathcal{F}$.

To this end, we make the choice $\mathcal{F}_0 = \int d^2 r f_0 $ with 
\begin{equation}
    f_0 = K|\nabla \mathbf{d}|^2 + (r_{\rm eff}+\Gamma_\perp)|\mathbf{d}_\perp|^2 + r_{\rm eff}|d_{z} - \sqrt{N}\psi |^2. 
\end{equation}
The parameters here are $r_{\rm eff}$ and $\psi$ which characterize the renormalized critical temperature and condensate respectively. 

We first evaluate the free energy of this ansatz.
To leading order in $N$ this is simply
\begin{equation}
    \mathcal{F}_0 = N T\int_{\bf q}\log\left(K\mathbf{q}^2 + r_{\rm eff} + \Gamma_\perp  \right) .
\end{equation}
This is independent of the condensate by construction.

We also need to calculate the various expectation values using this ansatz. 
We find the relevant correlation functions are 
\begin{equation}
    \langle d_z \rangle = \sqrt{N} \psi
\end{equation}
and 
\begin{equation}
    \langle |\mathbf{d}_\perp|^2 \rangle = N T\int_{\bf q}\frac{1}{K\mathbf{q}^2 + r_{\rm eff} + \Gamma_\perp} \equiv N I
\end{equation}
where $I$ is the integral, which gives the normal fluid component. 

We must also evaluate $\langle \Delta \mathcal{F}\rangle$ in this ansatz, which gives to leading order in $N$ 
\begin{equation}
    \langle \Delta \mathcal{F} \rangle = ( r - r_{\rm eff} ) \langle |\mathbf{d}_\perp|^2 \rangle + r \langle |d_z|^2\rangle + \frac{u+v}{2N}\langle |\mathbf{d}|^4 \rangle - \frac{v}{2N}\langle |\mathbf{d}^2 |^2 \rangle.
\end{equation}
The first two terms give $N(r - r_{\rm eff})I + N r |\psi|^2 $.
To leading order in $N$ the interactions give 
\begin{equation}
    \frac{1}{N}\langle |\mathbf{d}|^4 \rangle = N \left[ |\psi|^4 + 2|\psi|^2 I + I^2 \right]
\end{equation}
and 
\begin{equation}
    \frac{1}{N}\langle |\mathbf{d}^2|^2 \rangle = N |\psi|^4 .
\end{equation}
We then obtain the effective free energy functional of 
\begin{equation}
    \mathcal{F} = N \left[ T\int_{\bf q}\log\left(K\mathbf{q}^2 + r_{\rm eff} + \Gamma_\perp  \right) + (r - r_{\rm eff})I + (r + (u+v) I + \frac{u}{2} |\psi|^2 )|\psi|^2 + \frac{u+v}{2} I^2 \right]. 
\end{equation}

We now minimize over $\psi$ and $r_{\rm eff}$ to obtain the phase diagram.
This yields two equations 
\begin{subequations}
\begin{align}
    & \left[ r + (u+v)I + u |\psi|^2 \right]\psi = 0 \\
    & I + (r - r_{\rm eff})\frac{\partial I}{\partial r_{\rm eff}}  - I + (u+v)\frac{\partial I}{\partial r_{\rm eff}} |\psi|^2 + (u+v) I \frac{\partial I}{\partial r_{\rm eff}}= 0.
\end{align}
\end{subequations}
Since $I$ is in general dependent on $r_{\rm eff}$ this last equation can be simplified to read 
\begin{equation}
    r  + (u+v)( |\psi|^2 + I) = r_{\rm eff}. 
\end{equation}
We see the condensate density follows $(r_{\rm eff} - v|\psi|^2 )\psi = 0 $ and thus the transition to non-zero $|\psi|$ occurs when $r_{\rm eff} =0$.
This gives the phase boundary of 
\begin{equation}
    r + (u+v) I = 0 \Rightarrow r + \frac{u +v}{4\pi K} T\log(\frac{C}{\Gamma_\perp}) = 0.
\end{equation}
For simplicity, we neglect the temperature dependence of the fluctuations and set $T = T_{\rm BCS}$, which simplifies the system and is formally justified provided the renormalization is not to too low temperature.
Defining the Ginzburg-Levanyuk number as $u/(4 \pi K N(E_F)) T_{\rm BCS} = \frac12 \textrm{Gi}_{(2)} $ and taking $C/N(E_F) \sim (E_F/T_{\rm BCS})^2$ we find the renormalized critical temperature of 
\begin{equation}
    T_c = T_{\rm BCS} \left( \frac{\tilde{\Gamma}_\perp T_{\rm BCS}}{E_F} \right)^{\textrm{Gi}_{(2)}}.
\end{equation}
We have defined untiless pinning energy $\tilde{\Gamma}_\perp = \Gamma_\perp/N(E_F)$. 

This goes to zero as $\Gamma_\perp^{\textrm{Gi}_{(2)}}$ and for large $\Gamma_\perp$ will ultimately be expected to saturate to $T_{\rm BCS}$. 
Above the renormalized $T_c$ we have appreciable pair fluctuations but no condensate. 
We now include the possibility for $4e$ pairing in order to complete the picture of the phase diagram. 

\subsection{4$e$ Pairing}

In order to accommodate $4e$ pairing, which we expect to reside at small $\Gamma_\perp$, we must modify the ansatz we use to evaluate the correlation functions. 
In particular, we include an off-diagonal symmetric coupling via 
\begin{equation}
    f_0 = K|\nabla \mathbf{d}|^2 + (r_{\rm eff}+\Gamma_\perp)|\mathbf{d}_\perp|^2 + r_{\rm eff}|d_{z} - \sqrt{N}\psi |^2 - \frac12 \left[ \overline{\chi}( \mathbf{d}_\perp^2 + (d_z - \sqrt{N}\psi)^2 ) + \chi (\overline{\mathbf{d}}_\perp^2 + (\overline{d}_z -\sqrt{N}\overline{\psi})^2 )\right] . 
\end{equation}
The parameters here are $r_{\rm eff}$ and $\psi$ which characterize the renormalized critical temperature and condensate respectively. 
The free energy of this ansatz is 
\begin{equation}
    \mathcal{F}_0 = N \frac{T}{2}\int_{\bf q} \left[ \log\left(K\mathbf{q}^2 + r_{\rm eff} + \Gamma_\perp + |\chi|\right) +\log\left(K\mathbf{q}^2 + r_{\rm eff} + \Gamma_\perp - |\chi|\right) \right].
\end{equation}
We have condensate expectation value 
\begin{subequations}
\begin{align}
&   \langle d_z \rangle = \sqrt{N} \psi \\
&   \langle | \mathbf{d}_\perp|^2 \rangle = N I \\
&   \langle  \mathbf{d}_\perp^2 \rangle = N F \\
&   \langle  \overline{\mathbf{d}}_\perp^2 \rangle = N F^* .
\end{align}
\end{subequations}
The integrals are in turn given by 
\begin{subequations}
\begin{align}
&   I = \frac{T}{2} \int_{\bf q} \frac{1}{K\mathbf{q}^2 + r_{\rm eff} + \Gamma_\perp + |\chi|} + \frac{1}{K\mathbf{q}^2 + r_{\rm eff} + \Gamma_\perp - |\chi|}  = \frac{1}{4\pi K} \frac{T}{2}\left[\log\left(\frac{C}{r_{\rm eff} +\Gamma_\perp + |\chi|}\right) +\log\left(\frac{C}{r_{\rm eff} +\Gamma_\perp -|\chi|}\right)  \right]\\
&   F = -\frac{T}{2} \frac{\chi}{|\chi|}\int_{\bf q} \frac{1}{K\mathbf{q}^2 + r_{\rm eff} + \Gamma_\perp + |\chi|} - \frac{1}{K\mathbf{q}^2 + r_{\rm eff} + \Gamma_\perp - |\chi|} = \frac{T}{2} \frac{\chi}{4\pi K|\chi|} \log\left(\frac{r_{\rm eff} + \Gamma_\perp + |\chi|}{r_{\rm eff} + \Gamma_\perp - |\chi|} \right) .
\end{align}
\end{subequations} 

The evaluation of $\langle \Delta \mathcal{F}\rangle $ gives 
\begin{equation}
    \langle\Delta \mathcal{F} \rangle = ( r - r_{\rm eff} ) \langle |\mathbf{d}_\perp|^2 \rangle + r \langle |d_z|^2\rangle + \frac{u+v}{2N}\langle |\mathbf{d}|^4 \rangle - \frac{v}{2N}\langle |\mathbf{d}^2 |^2 \rangle + \frac12 \left[ \overline{\chi}\langle ( \mathbf{d}_\perp^2 + (d_z - \sqrt{N}\psi)^2 )\rangle + \chi \langle (\overline{\mathbf{d}}_\perp^2 + (\overline{d}_z -\sqrt{N}\overline{\psi})^2 )\rangle \right]. 
\end{equation}
The nonlinear terms are now evaluated as 
\begin{equation}
    \frac{1}{N}\langle |\mathbf{d}|^4 \rangle = N \left[ |\psi|^4 + 2|\psi|^2 I + I^2 \right]
\end{equation}
as before, and 
\begin{equation}
    \frac{1}{N}\langle |\mathbf{d}^2|^2 \rangle = N \left[ |\psi|^4 +  \psi^2 \overline{F} + F \overline{\psi}^2 + |F|^2 \right].
\end{equation}
We then get the free energy functional of 
\begin{multline}
    \frac{1}{N}\mathcal{F} =\frac12 T\int_{\bf q}\log\left(K\mathbf{q}^2 + r_{\rm eff} + \Gamma_\perp  + |\chi| \right) +\log\left(K\mathbf{q}^2 + r_{\rm eff} + \Gamma_\perp  - |\chi| \right) \\
    + (r - r_{\rm eff})I + (r + (u+v) I + \frac{u}{2} |\psi|^2 )|\psi|^2 + \frac{u+v}{2} I^2  - \frac{v}{2}\left(  \psi^2 \overline{F} + F \overline{\psi}^2 + |F|^2 \right)+\frac12( \overline{\chi}F + \overline{F}\chi ) . 
\end{multline}

The variational procedure gives a system of equations 
\begin{subequations}
\begin{align}
&   \left( r + (u+v)I+u|\psi|^2 \right)\psi - v F \overline{\psi}= 0  \\
&   r_{\rm eff} = r + (u+v) ( I + |\psi|^2 ) \\
&   \chi = v( \psi^2 + F) .
\end{align}
\end{subequations}
With the integrals in turns being functions of $\chi,\overline{\chi},r_{\rm eff}$. 

We have four cases; normal state, $2e$ condensate, $4e$ condensate, and coexistence. 
Starting from the normal state, where $\psi = \chi = 0$ we imagine lowering the temperature and find which phase we enter into first. 
For small $\Gamma$ we expect to first enter into the $4e$ condensate, which has a phase boundary determined by 
\begin{subequations}
\begin{align}
&   r_{\rm eff}- r= (u+v)I = (u+v) \frac{T}{4\pi K} \frac12 \left[ \log\left(\frac{C}{r_{\rm eff} + \Gamma_\perp + |\chi|}\right) + \log\left(\frac{C}{r_{\rm eff} + \Gamma_\perp -|\chi|}\right)\right]  \\
&   \chi = vF = v \frac{T}{2} \frac{\chi}{4\pi K|\chi|} \log\left(\frac{r_{\rm eff} + \Gamma_\perp + |\chi|}{r_{\rm eff} + \Gamma_\perp - |\chi|} \right) .
\end{align}
\end{subequations}
We again approximate these by replacing $T = T_{\rm BCS}$ and $\tilde{\chi} = |\chi|/N(E_F)$ to get 
\begin{equation}
2\tilde{\chi} =\textrm{Gi}_{(2)}  \textrm{artanh}\left(\frac{\tilde{\chi}}{\tilde{r}_{\rm eff} + \tilde{\Gamma}_\perp}\right).
\end{equation}
The first equation gives 
\begin{equation}
\tilde{r}_{\rm eff} = \tilde{r} +  \textrm{Gi}_{(2)} \log\left(\frac{C}{\sqrt{(\tilde{r}_{\rm eff} + \tilde{\Gamma}_\perp)^2 - \tilde{\chi}^2} }\right).
\end{equation}
We find the critical temperature of the transition by noting that a nontrivial solution for $\tilde{\chi}$ only appears once 
\begin{equation}
    \tilde{r}_{\rm eff}^\star = \frac12 \textrm{Gi}_{(2)}-\tilde{\Gamma}_\perp .
\end{equation}
At this point $\tilde{\chi} =0$ and we can solve for the physical temperature $T^\star = T_{\rm BCS} e^{\tilde{r}^\star}$ to get 
\begin{equation}
    \tilde{r}^\star = \frac12 \textrm{Gi}_{(2)} - \tilde{\Gamma}_\perp - \textrm{Gi}_{(2)} \log\left(2C/\textrm{Gi}_{(2)}\right).
\end{equation}
We see the critical temperature decays exponentially with the induced spin-orbit coupling. 

We see that the two phases boundaries coincide when 
\begin{equation}
    -\textrm{Gi}_{(2)} \log(C/\tilde{\Gamma}_\perp) = \frac12\textrm{Gi}_{(2)} - \tilde{\Gamma}_\perp - \textrm{Gi}_{(2)} \log\left(2C/\textrm{Gi}_{(2)}\right). 
\end{equation}
We find solution for the critical $\tilde{\Gamma}_c$ of 
\begin{equation}
    \tilde{\Gamma}_c +\textrm{Gi}_{(2)} \log(2\tilde{\Gamma}_c/\textrm{Gi}_{(2)}) = \frac12\textrm{Gi}_{(2)}  . 
\end{equation}
In fact, this equation can be solved easily to give $\tilde{\Gamma}_c = \frac12\textrm{Gi}_{(2)}$. 
We therefore will have an intersection of the two different condensate phases and we must consider their coexistence. 

At $\Gamma_\perp =0 $ and estimateing $1/\sqrt{C} = \textrm{Gi}_{(2)}\sim.1$ we find $T^*$ is of order $.5T_{\rm BCS}$ at zero spin-orbit coupling, which is remarkably large, atleast before the BKT physics is accounted for.

We start by considering the appearance of the $d_z$ condensate out of the $4e$ condensate, which proceeds by a $\mathbb{Z}_2$ spin-nematic transition associated to the phase of the $d_z$ condensate with respect to the phase of the quasi-long range ordered $4e$ condensate. 
We find the boundary of this transition by finding when the finite condensate appears. 
The relevant equation is 
\begin{equation}
    \left( r + (u+v)I+u|\psi|^2 \right)\psi - v F \overline{\psi}= 0 .
\end{equation}
We can take the $4e$ order parameter $F$ to be real, in which case we find the condensate locks into phase with $4e$ condensate, so as to minimize the Josephson coupling due to the $v$ nonlinearity.
Alternatively, we can recall that this nonlinearity essentially penalized the nonunitary fluctuations and therefore the lower energy state comes from the condensate being in phase so that the pairing remains unitary. 
Thus, the transition is spin nematic but preserves time-reversal symmetry, rather than the other case of a spin polarized state emerging which happens when the Josephson coupling between the $4e$ and $d_z$ condensates is negative. 

Armed with this, we then proceed to linearize this equation around small $\psi$ real. 
We get $ \left( r + (u+v)I \right)\psi - v F \psi = 0$.
This gives a critical point of 
\begin{equation}
  r_{\rm eff} - vF  =0 .
\end{equation}
Here we have noted that $r_{\rm eff} = r+ (u+v)I$.
The integrals in turn require solving 
\begin{equation}
    \chi = r_{\rm eff}. 
\end{equation}
Together these give equation 
\begin{equation}
\tilde{r}_{\rm eff} = \frac14\textrm{Gi}_{(2)}\log(\frac{2 \tilde{r}_{\rm eff} + \tilde{\Gamma}_\perp}{\tilde{\Gamma}_\perp}) .
\end{equation}
This always has a solution for positive $r_{\rm eff}$ provided $\textrm{Gi}_{(2)}/2\tilde{\Gamma}_\perp > 1$. 
At the point where $\tilde{\Gamma}_{\perp} = \frac12 \textrm{Gi}_{(2)} $ we see that $\tilde{r}_{\rm eff}$ goes to zero and thus $\tilde{\chi}$ goes to zero, indicating indeed this curve terminates at $\tilde{\Gamma}_c$ where the $2e$ condensate takes over from the $4e$ condensate. 
We therefore see that there is a second transition which occurs within the $4e$ condensed phase characterized by the spin nematic transition, and this is a continuous second-order transition persisting up to the point where the $4e$ condensate disappears in a lambda point.

\section{Spin-Susceptibility}
We now briefly consider the susceptibility of the condensate to an in-plane magnetic field in the pseudogap region, which may serve as an experimental signature to help confirm this model. 
The spin-susceptibility is evaluated as 
\begin{equation}
    \chi_{\parallel} = -\frac{\delta^2\mathcal{F}}{\delta\mathbf{H_{\parallel}}^2}
\end{equation}
where the coupling to magnetic field is governed by the coupling $\mathbf{H}_{\parallel}\cdot(i\overline{\mathbf{d}}\times\mathbf{d})$, and evaluated at zero external field. 
We thus get 
\begin{equation}
    \chi_{\parallel} = \frac{\mu_{\rm eff}^2}{T} \langle (i\overline{\mathbf{d}}\times\mathbf{d})(x) (i\overline{\mathbf{d}}\times\mathbf{d})(x') \rangle.
\end{equation}
The correlation functions are evaluated in the disordered phase and to leading order in $N$ are given by (ignoring the $4e$ condensate)
\begin{equation}
    \langle {d}_\alpha(x) \overline{{d}}_\beta(x')\rangle = -\int_{\bf q} e^{i\mathbf{q}\cdot(\mathbf{x-x'})}\frac{\mu_{\rm eff}^2T}{K\mathbf{q}^2 + r_{\rm eff} + \Gamma_\perp}\delta_{\alpha\beta}. 
\end{equation}
We then get by Wick's theorem 
\begin{equation}
    \chi_{\parallel}^{ml} = -\frac{\mu_{\rm eff}^2}{T} \epsilon^{abm}\epsilon^{jkl} \langle \overline{d}_a(x) d_k(x')\rangle \langle d_b(x)\overline{d}_j(x')\rangle.
\end{equation}
Contracting indices and picking out the dominant term in $1/N$ we find for the response to homogeneous field 
\begin{equation}
    \chi_{\parallel}^{lm}(\mathbf{Q}\to 0) = \delta_{lm}N \mu_{\rm eff}^2T \int_{\bf q} \left(\frac{1}{K\mathbf{q}^2 + r_{\rm eff} +\Gamma_\perp}\right)^2 .
\end{equation}
This integral gives the fluctuation contribution of 
\begin{equation}
    \chi_{\parallel}^{lm}(\mathbf{Q}\to 0) = \delta_{lm}\frac{N\mu_{\rm eff}^2T}{4\pi K} \frac{1}{r_{\rm eff} +\Gamma_\perp} .
\end{equation}
We find that due to the superconducting spin fluctuations above $T_c$ we find a divergent paramagnetic response, similar to a Curie-Weiss behavior.
This contribution will be proportional to $\textrm{Gi}_{(2)}$ and is cutoff for finite spin-orbit coupling at the critical point as
\begin{equation}
    \chi_{\parallel} \sim \frac{\textrm{Gi}_{(2)}}{\tilde{\Gamma}_\perp} ,
\end{equation}
while the divergence at zero spin-orbit coupling has form 
\begin{equation}
    \chi_{\parallel} \sim \frac{\textrm{Gi}_{(2)}}{r_{\rm eff}}. 
\end{equation}
This will have the same characteristic critical exponent as the one associated with $r_{\rm eff}$, which also governs the order parameter susceptibility.  
We note that this change in the spin susceptibility is purely due to the soft triplet order parameter fluctuations, as evidenced by it being proportional to $\textrm{Gi}_{(2)}$, and in a mean-field transition into a triplet phase it is expected to vanish. 
For simplicity, we have neglected the role of the $4e$ pairing although it would be interesting to consider them in the future since it may have a competing effect since it is essentially spin-singlet.
This may then manifest as an apparent divergence of the spin susceptibility which abruptly turns over into a singlet-like response at $T^*$.
Depending on the strength of fluctuations, this apparent initial upturn may still be visible and would still be anomalous in comparison to a BCS-like singlet superconductor.

\section{Estimates of density of states and the Fermi energy}
In this section, we discuss our procedure for estimating the density of states at the Fermi level and the Fermi energy that are required for calculating the Ginzburg-Landau parameters. 

\begin{figure}
    \includegraphics[scale=1]{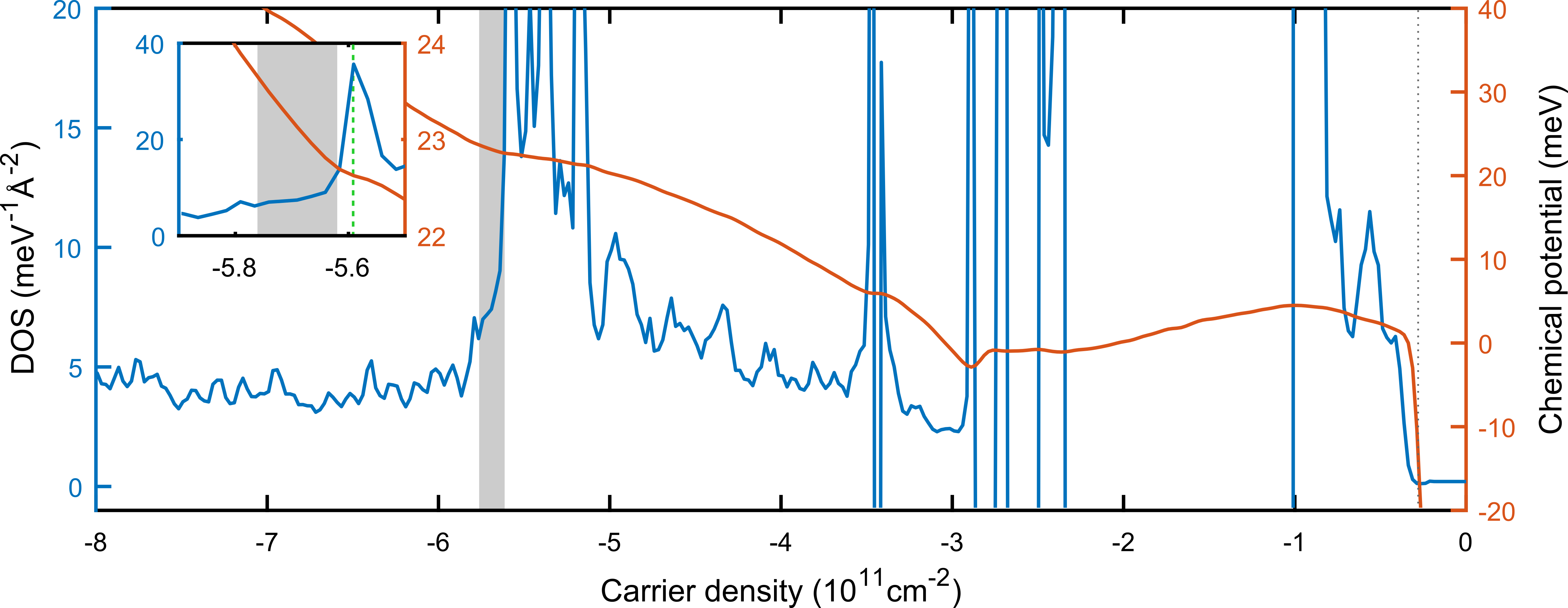}
    \centering
    \caption{
    Density of states and chemical potential for Bernal bilayer graphene Extracted from Ref.~\cite{Zhou.2022k0q}.. The black dotted line marks the edge of the valence band of BBG in the presence of electric field. The grey shaded region denotes the density range in which spin triplet superconductivity is observed under small in-plane magnetic field. The inset shows a zoom in near the grey shaded region. The green dotted line marks the isospin phase transition near the superconducting dome edge.
    }
    \label{fig:dos}
\end{figure}

In Ref\cite{Zhou.2022k0q}, inverse compressibility measurements were used to identify the broken isospin phases in BBG at higher temperatures before the spin-triplet superconductivity emerges. Inverting the measured inverse compressibility data directly yields the many-body density of states in the presence of strong correlation. The blue trace in Fig. \ref{fig:dos} shows the density of state at $D=1.02$V/nm obtained from Ref.\cite{Zhou.2022k0q}. The grey shaded region marks the density range within which superconductivity is observed. 
We therefore estimate the density of state to be 7 $\textrm{meV}^{-1}\si{\angstrom}^{-2}$.

Integrating the same inverse compressibility data allows us to extract the electronic chemical potential as shown in Fig.~\ref{fig:dos} (red trace). The black dotted line denotes the valence band edge and thus the origin of the Fermi energy. The chemical potential reaches approximately 23 meV at the optimal doping. To interpret the extracted chemical potential, we note that the system undergoes a cascade of isospin phase transitions that lead to one large and one small Fermi surfaces at the optimal doping \cite{Zhou.2022k0q,DelaBarrera2022,Zhang.2022}. These two Fermi surfaces are each two-fold degenerate and characterized by two different Fermi energies. Although it is challenging to precisely disentangle their contributions to the change in chemical potential, one rough estimate is to assume that the large Fermi surface dominates the total change in chemical potential counting from the charge neutrality, whereas the small Fermi surface is mainly responsible for the changes counting from the phase transition point (see green dotted line in the inset of Fig. \ref{fig:dos}). Under these assumptions, we find that the Fermi energy associated with the small Fermi surface is roughly 0.6 meV. Since the system has gone through already one phase transition that resets the chemical potential, we estimate the Fermi energy associated with the large Fermi surface to be roughly half of the total change in chemical potential, that is approximately 10 meV. While we do not know which Fermi surface is more responsible for the spin-triplet pairing, we note that Pauli-limit violation is also observed in twisted graphene multilayers\cite{Cao2021,Park2022} and the Fermi energies in these systems ($\sim$ 1 meV) are very similar to that of the small Fermi surface in BBG. Since our theory is motivated solely based on general symmetry arguments and is broadly applicable to all graphene-based spin-triplet superconductors, we use 0.6 meV for the Fermi energy in our calculations.

\end{document}